\documentclass[twocolumn]{aastex631}

\newcommand\pnbd{BD+30$^{\circ}$3639}

\shorttitle{Median Energy Imaging}
\shortauthors{Montez}

\graphicspath{{./}{figures/}}
\begin{document}

\title{Discovery of Radial Spectral Hardening in the Hot Bubble of Planetary Nebula \pnbd{} with Median Energy Imaging}


\author[0000-0002-6752-2909]{Rodolfo Montez Jr.}
\affiliation{Center for Astrophysics $\vert$\ Harvard\ \&\ Smithsonian, Cambridge, MA, USA}
\correspondingauthor{Rodolfo Montez Jr.}
\email{rodolfo.montez.jr@gmail.com}

\begin{abstract}
We introduce a new imaging analysis technique to study the spatial distribution of the X-ray emission from the hot bubble of planetary nebula \pnbd{}. 
Hot bubble emission is typically photon-starved, thus limiting the methods for spatial-spectral analysis, however, this new technique uses the statistics of photon energies across the nebula to identify spatial variations. 
Using the median energy value of the X-ray photons, we identified a rise in median energy values towards the projected edge of the nebula, which we refer to as radial spectral hardening. 
We explored the origin of this radial spectral hardening with X-ray spectral analysis of distinct regions of high- and low-median energy values.
Given that the hot bubble is embedded within a young, dense, planetary nebula, we argue that the radial spectral hardening is due to an increased column density at the projected nebular edge. 
Median energy imaging provides a promising new methodology for exploring the spatial variations in faint extended X-ray sources. 
\end{abstract}


\section{Introduction} \label{sec:intro}

A planetary nebula (PN) forms during the late evolutionary stages of low- to intermediate-mass star (1-8~$M_{\odot}$) when a nascent fast stellar wind collides with the older, denser, and slower asymptotic giant branch (AGB) stellar wind \citep[][]{1978ApJ...219L.125K}. 
According to this interacting stellar wind theory, material from the AGB wind is swept into the nebular shell and ionized by the hot central star. 
This wind-wind interaction generates a shocked region that reaches X-ray emitting temperatures ($>10^{6}~{\rm K}$) and is called the ``hot bubble''.
Hot bubbles are expected to scale with the square of the fast wind velocity, however, {\it Chandra} and {\it XMM-Newton} observations indicate that plasma temperatures are much cooler than expected \citep[e.g.,][]{2008ApJ...672..957K,2013ApJ...767...35R}. 
Understanding the origin of these cooler plasma temperatures is a unresolved area of research being explored with a variety of approaches \citep{2003ApJ...583..368S,2007MNRAS.375..137A,2008A&A...489..173S,2018A&A...620A..98H,2016MNRAS.463.4438T,2018MNRAS.478.1218T}. 

The {\it Chandra} X-ray Observatory unambiguously resolved hot bubble emission for the first time with observations of the PN \pnbd{}. The plasma temperature was found to be $3\times10^{6} {\rm ~K}$ by studying the low energy resolution CCD X-ray spectrum \citep{2000ApJ...545L..57K}. 
High-resolution grating X-ray spectroscopy revealed that the hot bubble emission was carbon-rich and iron-deficient and provided evidence for a non-isothermal plasma with plasma temperatures ranging from 1.7 to $2.9\times10^{6}~{\rm K}$ \citep{2009ApJ...690..440Y}. 
Additionally, hot bubbles are embedded within the optically- and infrared-bright nebular shells and variations of this overlying nebular material are believed to influence the detected X-ray emission \citep{2002ApJ...581.1225K,2003ApJ...589..439M}. 
The spatial and spectral decomposition of the X-ray emission from the hot bubble in NGC 7027 \citep{2018ApJ...861...45M} qualitatively demonstrated the degree of influence that the surrounding nebular material can have on the observed hot bubble plasma. 
The spatial resolution of {\it Chandra} is sufficient for the mapping planetary nebulae hot bubbles, but most hot bubbles studied by {\it Chandra} are photon-starved \citep{2012AJ....144...58K,2014ApJ...794...99F}. 
In this study, we explore a novel methodology designed to glean spatial information across the hot bubble of \pnbd{} using the photon statistics of the X-ray emission.

\section{Observations} \label{sec:obs} 

\subsection{{\it Chandra} Observations}

\pnbd{} has been observed for a total of nearly 400 ks with {\it Chandra} ACIS-S imaging and zeroth-order ACIS-S gratings observations.
Although it is tempting to compile all of the observations, the time-dependent sensitivity due to build-up of a contaminant on the ACIS optical blocking filter \citep{2004SPIE.5165..497M} would limit the utility of the methodology employed in this work. 
This is especially problematic because \pnbd{} primarily emits soft photons ($\lesssim 1$~keV), where the most change in detector sensitivity has occurred.
Instead, we focus only on the observations acquired in Cycle 9 (ObsIDs 9932 and 10821 with exposures of 38.1 and 39.1 ks and acquired on 2009-01-27 and 2009-01-22, respectively; PI: Yu). 
Instrumental changes in sensitivity are negligible since these two observations occurred within five days of each other.

We analyzed the observations with CIAO \citep[version 4.11][]{2006SPIE.6270E..60F} and the calibration database CALDB 4.8.3. 
The \texttt{chandra\_repro} CIAO script was used to reprocess the data with sub-pixel event repositioning \citep{2003ApJ...590..586L,2004ApJ...610.1204L} and the \texttt{merge\_obs} CIAO script was used to merge the two observations. 
The 77 ks merged observation yields 9485 counts in the 0.15 to 1.5 keV energy range from a region consistent with the nebula (an ellipse with a semi-major and semi-minor radii of $3^{\prime\prime}$ and $2.5^{\prime\prime}$, respectively). 
The average count density is $\approx 400$ counts per square arcsecond or $\approx 100$ counts per ACIS-S native pixel ($0\farcs492\times0\farcs492$).

We have also extracted source and background spectra from ObsID 10821 with the CIAO script \texttt{specextract} following the analysis thread for extended sources\footnote{CIAO analysis thread is located at: \url{https://cxc.cfa.harvard.edu/ciao/threads/extended/}}. 
We modeled the emission using Astrophysical Plasma Emission Code (APEC) models with variable abundances \citep{2001ApJ...556L..91S} and the Tuebingen-Boulder ISM absorption model \citep{2000ApJ...542..914W}. 
In \S\ref{sec:discussion} we provide further details on the spectral analysis. 

\subsection{{\it Hubble} Observations} 

We obtained the H$\alpha$ $\lambda$656.3 nm narrow-band images of \pnbd{} from the {\it Hubble} Space Telescope. 
The H$\alpha$ narrow-band filter (F656N) image from {\it Hubble} proposal 11122 (PI: Balick) was acquired from the {\it Hubble} Legacy Archive\footnote{https://hla.stsci.edu/}. 
These observations were obtained on 2008-03-02, within a year of the Cycle 9 {\it Chandra} observations. 
The F656N images were flux scaled according to \citet{1997hstc.work..338D}, however, we did not correct for contamination from [NII] (F658N), nor the continuum from the F547M filter (not available). 
The optical image is dominated by H$\alpha$ emission. 
We performed a cosmetic cleaning of hot pixels, saturation regions, and cosmic rays with negligible impact on our analysis. 
The central star has been masked for display purposes. 

\begin{figure*}[ht]
\centering
\includegraphics[width=6.5in]{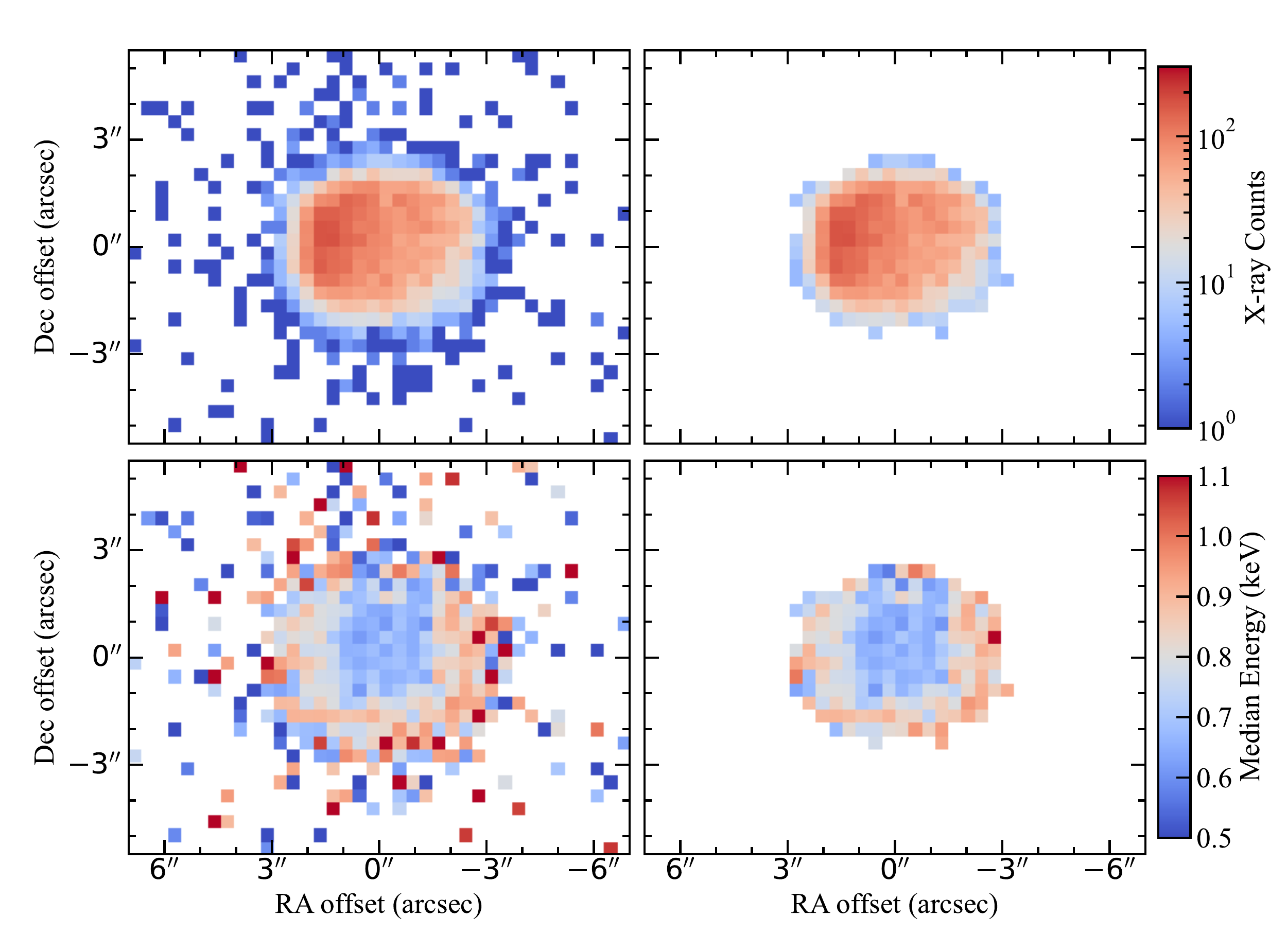} 
\caption{Count ({\it top row}) and median energy ({\it bottom row}) images of \pnbd{} derived from the Cycle 9 observations obtained by {\it Chandra}. All images have bins comparable to the native ACIS-S pixel size ($0\farcs492\times0\farcs492$). The images in the {\it left column} depict the full X-ray observation, while the images in the {\it right column} are filtered to only show pixels with 5 or more counts. The median energy image is derived from the energy distributions in each count image pixel.}
\label{fig:bd30_imaging}
\end{figure*}

\section{Median Energy Image} \label{sec:analysis} 

\begin{figure*}[ht]
\centering
\includegraphics[width=6.5in]{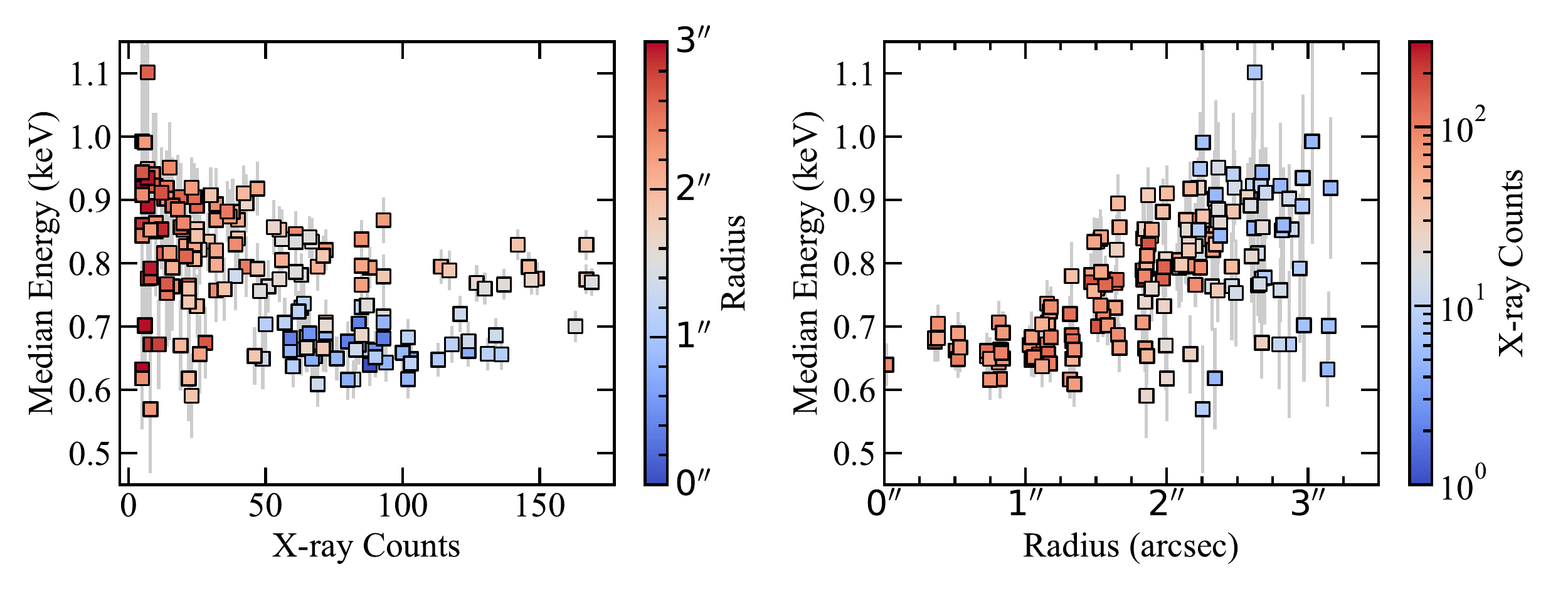} 
\caption{Pixel values from the count and median energy images. In the {\it left} panel, the median energy pixel values are plotted as a function of the number of X-ray counts and the symbols are color-coded by the radial distance from the central star. In the {\it right} panel, the radial profile of the median energy pixels is plotted and the symbols are color-coded by the number of X-ray counts. In each panel, the error bars are standard errors on median energy values as described in the text and we only plot pixels with at least 5 counts.  }
\label{fig:bd30_pixels}
\end{figure*}

We introduce an imaging product called the median energy image, which consists of an image wherein each pixel is generated from the statistical properties of photons that reside within the pixel boundaries. 
We generate a number of evenly distributed spatial bins based on the desired pixel size, $w$, and overall image size. 
Each photon is collected into its appropriate bin based on its physical coordinates and we determine the median value of the photon energies, $\tilde{E}$, the standard deviation of the photon energies, $\sigma_{E}$, and the total number of photons, $N$, for the collection of photons in each pixel. 

To estimate the variance of the median value of the photon energies, $\sigma_{\tilde{E}}^{2}$, we adopt the standard error on the median,
\begin{equation}
\sigma_{\tilde{E}}^{2} = \frac{\pi}{2} \frac{\sigma_{E}^2}{N}
\end{equation}
where $N$ is the number of photons in a given pixel. 
We note that this formula is intended for normal distributions and typical hot bubble plasma emission is not normally distributed. 
As an alternative, we considered a boot-strapping error estimation derived from thousands of draws of $N$ distinct photons from the observed plasma distribution. 
We found that the standard error on the median provided realistic uncertainty estimates and boot-strapping typically resulted in overly optimistic error estimates (smaller errors).

The median values of the photon energies, hereafter referred to as the ``median energy'' values, $\tilde{E}$, median energy error values, $\sigma_{\tilde{E}}$, and total count values, $N$, are used to populate the pixels in 2D arrays that form the median energy image, median energy error image, and the count image, respectively. 
In Figure~\ref{fig:bd30_imaging}, we depict the count and the median energy images. 

We calculated the radius from the central star of \pnbd{} to each pixel. 
These radii are color-coded in the left panel of Figure~\ref{fig:bd30_pixels} in the count-median energy plane. 
In the right panel of Figure~\ref{fig:bd30_pixels}, we plot the radial distribution of the pixels in the median energy images color-coded by the number of X-ray counts in each pixel. 
Error bars for the median energy values are included in both panels based on $\sigma_{\tilde{E}}$. 
The radial extent of the X-ray emission from \pnbd{} is $\lesssim 3\farcs0$ (see Figure~\ref{fig:bd30_imaging}).

\begin{figure*}[ht]
\centering
\includegraphics[width=6.5in]{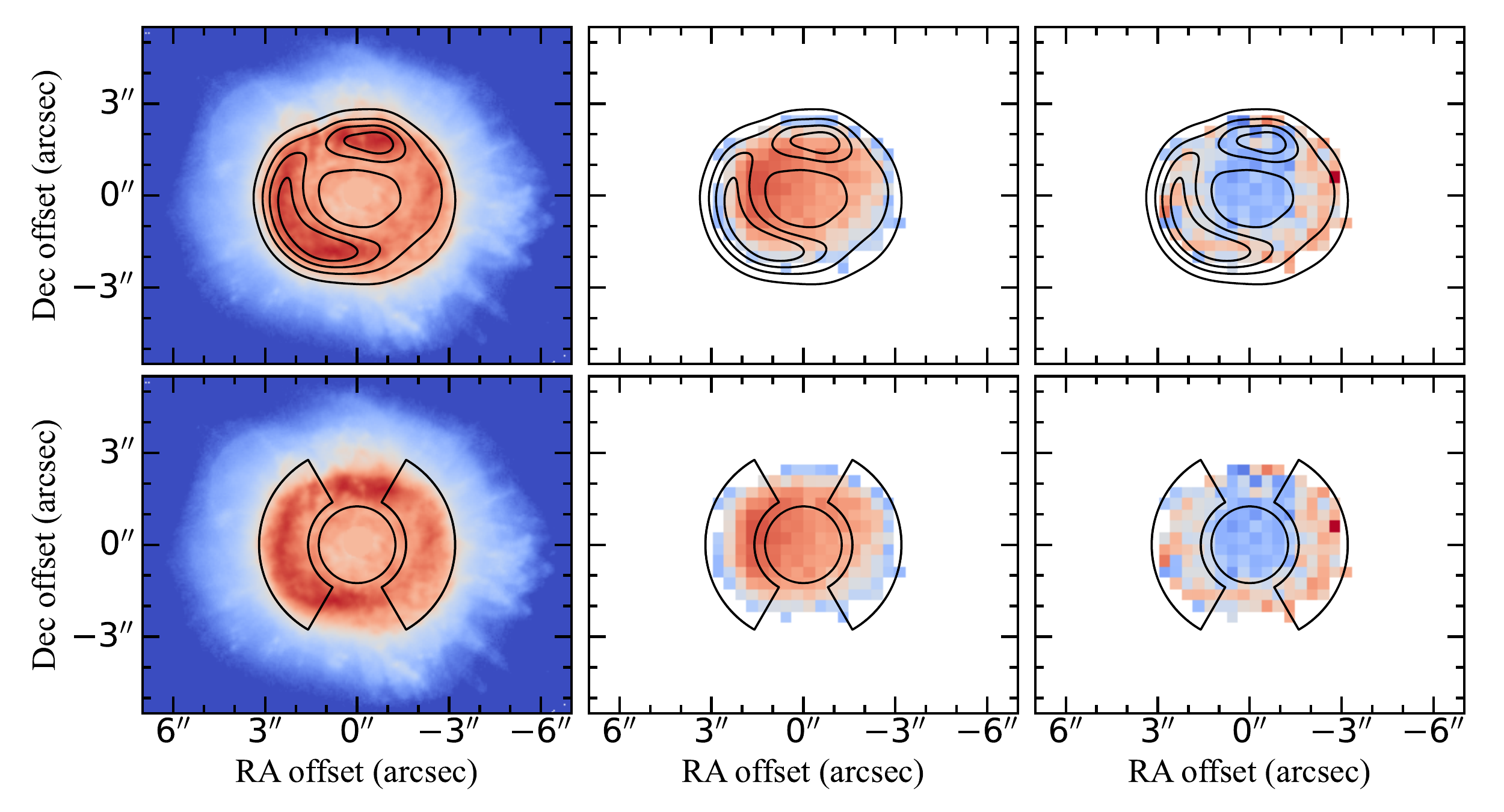}
\caption{From {\it left} to {\it right} we depict the {\it Hubble} H$\alpha$ image, the {\it Chandra} X-ray count image, and the median energy image. In the {\it top row} we overlay contours from the smoothed {\it Hubble} image and in the {\it bottom row} we overlay spectral extraction regions from the high-median energy region (outer wedge-shaped regions) and a low-median energy region (inner circular region).  }
\label{fig:bd30_multi}
\end{figure*}

In Figure~\ref{fig:bd30_multi}, we depict the {\it Hubble} H$\alpha$ emission, the X-ray count, and the median energy images. 
In the top row, contours from the smoothed H$\alpha$ image are overlaid on all the images, while in the bottom row the spectral extraction regions discussed in \S\ref{sec:discussion} are overlaid on the images. 
The central star in the H$\alpha$ image is masked and filled in with the local average flux. 
The X-ray count and median energy images are filtered to only show those pixels with at least 5 counts.

\section{Results} \label{sec:results} 

The median energy imaging of \pnbd{} reveals a distinct morphology compared to the morphology seen in the X-ray count image. 
The median energy image exhibits a partial ring-like structure of higher median energies ($\tilde{E} \sim0.8-0.9$~keV) and a central region with lower median energies ($\tilde{E}\sim0.6-0.7$~keV). 
Based on Figure~\ref{fig:bd30_pixels}, there is no apparent trend between the number of counts and the median energy values. 
Specifically, higher median energy values exist for both low and high count pixels. 
The partial ring-like structure of higher median energy is also apparent in the radial profile shown in Figure~\ref{fig:bd30_pixels}, where the rise in median energy begins at $\approx 1\farcs0$ and rises till the edge of the nebula at $\approx 2\farcs5$ to $3^{\prime\prime}$. 
We refer to this apparent spectral hardening of the hot bubble emission as ``radial spectral hardening'' and now discuss its possible origins.

\begin{deluxetable}{lrrr}
\tablecaption{Spectral fits for median energy-defined regions.\label{tab:specfits}}
\tablecolumns{4}
\tablehead{\colhead{Parameter} & \colhead{Model I} & \colhead{Model II} & \colhead{Model III} 
}
\startdata
$\chi_{\rm reduced}^2$ & 1.631 & 1.236 & 1.211 \\ 
Degrees of Freedom & 111 & 114 & 114 \\ 
\sidehead{High median energy region (R1):}
$N_{\rm H,R1}\ (10^{22}{\rm ~cm}^{-2})$ & 0.25$^{+0.02}_{-0.02}$ & 0.25$^{+0.03}_{-0.03}$ & 0.31$^{+0.04}_{-0.04}$ \\ 
$T_{\rm X,R1}$ (K) & 2.26$^{+0.05}_{-0.05}$ & 2.40$^{+0.15}_{-0.15}$ & 2.19$^{+0.10}_{-0.10}$ \\ 
O$_{\rm R1}$ & 0.23$^{+0.08}_{-0.08}$ & 0.44 & 0.44 \\ 
Ne$_{\rm R1}$ & 1.7$^{+0.4}_{-0.4}$ & 2.03 & 2.03 \\ 
norm$_{\rm R1}\ (10^{-4})$ & 13.1$^{+2.6}_{-2.6}$ & 9.5$^{+2.2}_{-2.2}$ & 14.3$^{+3.5}_{-3.5}$ \\
\sidehead{Low median energy region (R2):}
$N_{\rm H,R2}\ (10^{22}{\rm ~cm}^{-2})$ & $=N_{\rm H,R1}$ & $=N_{\rm H,R1}$ & 0.15$^{+0.03}_{-0.03}$ \\ 
$T_{\rm X,R2}$ (K) & $=T_{\rm X,R1}$ & 1.80$^{+0.10}_{-0.10}$ & $=T_{\rm X,R1}$ \\ 
O$_{\rm R2}$ & 0.21$^{+0.08}_{-0.08}$ & 0.44 & 0.44 \\ 
Ne$_{\rm R2}$ & 0.8$^{+0.3}_{-0.3}$ & 2.03 & 2.03 \\ 
norm$_{\rm R2}\ (10^{-4})$ & 9.5$^{+1.4}_{-1.4}$ & 10.6$^{+2.8}_{-2.8}$ & 4.75$^{+1.0}_{-1.0}$ \\
\enddata
\tablecomments{\footnotesize Model descriptions: \\{\it Model I}: oxygen (O) and neon (Ne) abundances allowed to vary in two regions independently. Column densities and temperatures are tied together ($N_{\rm H,R2} = N_{\rm H,R1}$ and $T_{\rm X,R2} = T_{\rm X,R1}$, respectively) and allowed to vary. \\
{\it Model II}: Abundances fixed, column densities are tied together ($N_{\rm H,R2} = N_{\rm H,R1}$ and allowed to vary, and temperatures are allowed to vary in two regions independently. \\
{\it Model III}: Abundances fixed, temperatures are tied together ($T_{\rm X,R2} = T_{\rm X,R1}$) and allowed to vary, and column densities are allowed to vary in two regions independently. \\
Errors are derived from 90\% confidence ranges. 
The abundances of carbon (C), nitrogen (N), magnesium (Mg), and iron (Fe) are set to those reported in \citet{2009ApJ...690..440Y}, 19.5, 0.17, 0.6, and 0.13, respectively, as are O, and Ne for Models II and III. 
See Figure~\ref{fig:bd30_multi} for depiction of regions and Figure~\ref{fig:bd30_spectra} for plots of the best-fit spectra. }
\end{deluxetable}

\section{Discussion} \label{sec:discussion}

The median value is known as a robust statistic, meaning that the median value provides a good representation of the distribution even in the presence of strong outliers and non-optimal distributions (e.g., non-normal and/or low count distributions).  
The utility of median energy values have been shown in various studies.
For example, in the {\it Chandra} Orion Ultradeep Project \citep{2005ApJS..160..379F}, the median energy values are used to estimate $N_H$ for point sources with a few to tens of counts. 
In \citet{2004ApJ...614..508H}, the median and quartile energy values are used to classify spectral properties of X-ray point sources with limited statistics. 
In the lower-temperature regime, the {\it Chandra} Planetary Nebula Survey (ChanPlaNS) used the median energy values and proxies for the column densities to characterize both point sources and extended sources of X-ray emission from PNe and their central stars \citep{2012AJ....144...58K,2014ApJ...794...99F}. 
Our analysis presents the first attempt to use the median energy values as an imaging tool. 

For well-behaved and well-characterized instrumental sensitivity and calibration, changes in the median energy values for a source modeled by an absorbed thermal plasma could be caused by variations in the elemental abundances, plasma temperature, and/or absorbing column. 
These three potential origins are not necessarily mutually-exclusive and have a degree of interdependence. 
As the plasma temperature changes so does the ionization stage and thus the strength of specific emission lines in the X-ray spectrum. 
Emission lines at lower energy values are more sensitive to absorption than those at higher energy values, so photons from C{\sc vi} at $\approx$0.37~keV will be absorbed more readily than Ne{\sc ix} at $\approx$0.92 keV. 
Additionally, absorption becomes less efficient as the plasma temperature rises since a hotter plasma emits fewer soft photons.
We independently consider each of these three potential origins in the case of \pnbd{} by modeling X-ray spectra from the high median energy and low median energy regions detailed in Figure~\ref{fig:bd30_multi}. 

\subsection{Varying Elemental Abundances} 

The elemental abundances of hot bubbles in PNe is an open question. 
The central star often has a distinct chemical composition compared to the nebular chemical composition. 
When modeling the hot bubble emission, one can assume the chemical composition of the central star or the nebula. 
However, when there is sufficient signal to constrain hot bubble abundances, the chemical composition is often found to be discrepant with both the central star and nebular abundances \citep[e.g.,][]{2019ApJ...886...30T}. 
Mixing between central star and nebular abundances are included in the theoretical efforts to understand hot bubble X-ray emission \citep{2008A&A...489..173S,2018A&A...620A..98H,2016MNRAS.463.4438T,2018MNRAS.478.1218T}.
In the X-ray emitting temperature ranges of hot bubbles, the abundances of carbon, nitrogen, oxygen, neon, and iron have the strongest influence on the spectral shape of the X-ray emission emission. 
At CCD spectral resolution, it is often difficult to obtain meaningful constraints on many of these elements, especially for carbon and iron \citep[see][for a detailed discussion]{2018ApJ...861...45M}.
However, the high energy resolution grating spectrum of \pnbd{}, which resolves the major ionized lines responsible for the X-ray emission, provides strong constraints on the important elemental abundances \citep{2009ApJ...690..440Y}.

\begin{figure*}[ht]
\centering
\includegraphics[width=6.5in]{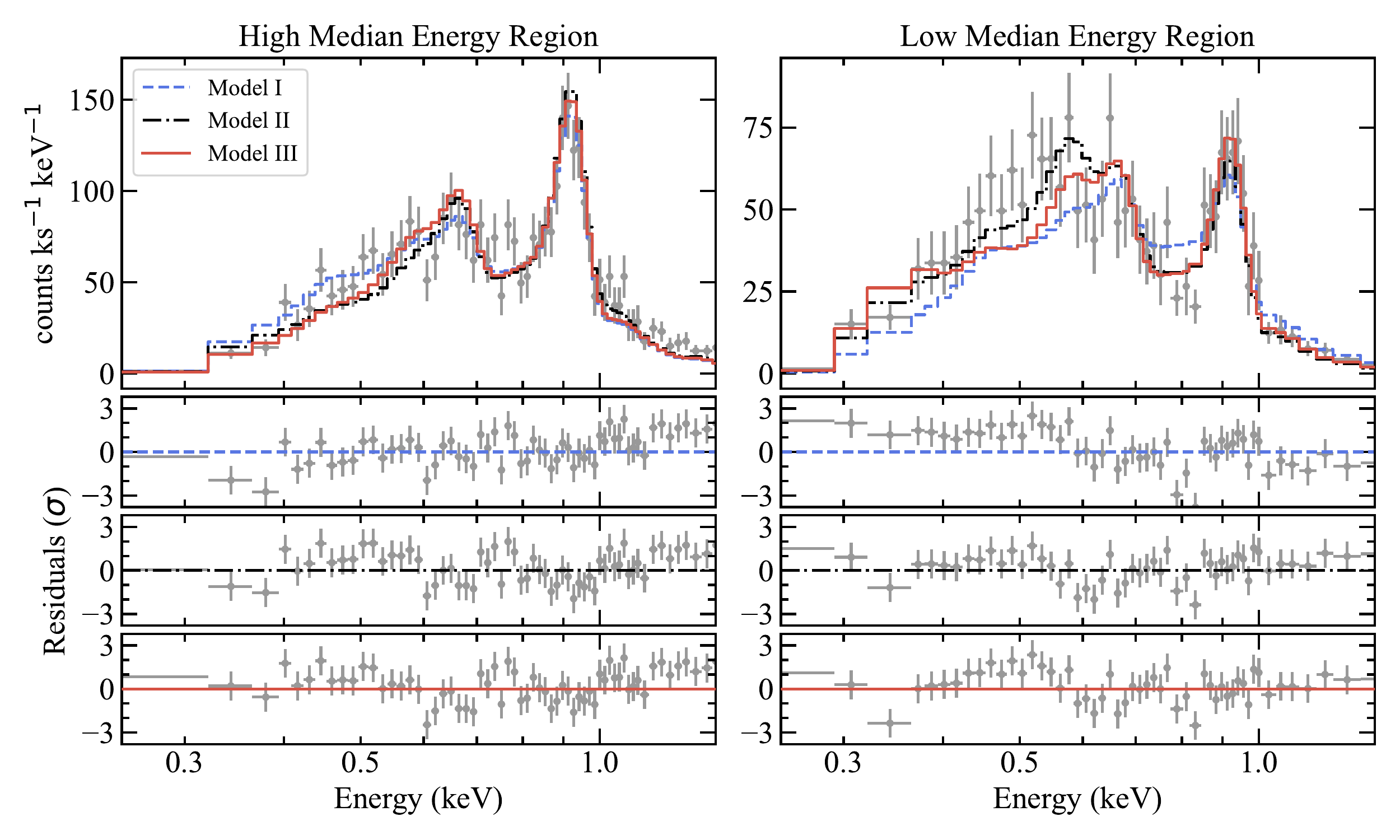}
\caption{X-ray spectra and spectral fits from the high ({\it left}) and low ({\it right}) median energy regions. There are three models fit to the spectra. These three models are described in \S\ref{sec:discussion} and Table~\ref{tab:specfits}. The three smaller panels in each row show the residuals produced by each model fit. }
\label{fig:bd30_spectra}
\end{figure*}

The apparent radial spectral hardening in \pnbd{} could be explained if the neon abundance increased radially outward (and/or if oxygen decreased radially inward).
To test this scenario, we assumed a single absorbing column and plasma temperature (fit as free parameters) and allowed the abundances for oxygen and neon independently vary in the two regions (Model~I in Table~\ref{tab:specfits}). 
As expected, the neon abundance was higher in the high-median energy region and lower in the low-median energy region, while oxygen trended towards a value that is a factor of two lower than that reported by \citet{2009ApJ...690..440Y}. 
In Figure~\ref{fig:bd30_spectra} we plot the spectra, best-fit model, and residuals for the low and high median energy regions. 
The fit to the spectrum of the low-median energy region is poor for energies $\lesssim0.5$~keV. 
We also attempted spectral fits to the carbon, nitrogen, and iron abundances but the abundances and plasma properties became unconstrained, suggesting that the spectrum is relatively insensitive to these parameters. 
Instead, we fixed the abundances of these elements to the single-temperature model in \citet{2009ApJ...690..440Y}\footnote{\citet{2009ApJ...690..440Y} do not detect nitrogen lines and argued that the non-detection places a strong upper limit on the nitrogen abundance, which is the value used in their spectral modeling.}. 

\subsection{Varying the Plasma Temperatures} 

Assuming a constant column density and a homogeneous chemical composition in the hot bubble, the radial spectral hardening could suggest a radially-increasing temperature. 
To explore this scenario, we fit a single column density to the low- and high-median energy regions with abundance values fixed to that from \citet{2009ApJ...690..440Y} and allowed two independent temperatures, one for each of the two regions (Model II in Table~\ref{tab:specfits}). 
As expected, the high-median energy region results in a higher plasma temperature ($T_{\rm X}=2.4\pm0.2 {\rm ~MK}$) compared to the low-median energy region ($1.8\pm0.1 {\rm ~MK}$).
For comparison, the spectral fits of the high-resolution grating X-ray spectroscopy of \pnbd{} led \citet{2009ApJ...690..440Y} to argue for a two-component model with temperatures of 1.7 and $2.9\times10^{6}~{\rm K}$. 
The spectral fit for this scenario is well-behaved and its goodness of fit is acceptable and better than that of Model I. 
It is worthwhile to note that heat conduction and nebular mixing studies predict radially-decreasing temperatures, in contrast to the prediction of Model II of radially-increasing temperatures. 

\subsection{Varying the Absorbing Columns}

For the cooler plasma temperatures measured from hot bubbles, intervening material can be efficient at absorbing the soft ($<1~{\rm keV}$) X-ray photons. 
Assuming the hot bubble is an isothermal plasma with homogeneous chemical composition, we performed spectral fitting on the two median energy regions assuming a single plasma temperature and two independent column densities (see Model III in Table~\ref{tab:specfits}).  
Similar to Model II, the spectral fit for this scenario is well-behaved and its goodness of fit is acceptable and better than that of Model I. 
The best fit parameters suggest that the column density in the high-median energy region is a factor of two higher than that in the low-median energy region. 

Since we view the hot bubble through the nebula, line of sights near the projected nebular edge will pass through more nebular material than those in the center, so higher column densities of nebular material at projected edge of nebula should be expected. 
The radius of the nebula projected onto the sky is 0.02 pc \citep{2012AJ....144...58K} and the thickness of the nebular shell is likely an order of magnitude smaller, so the ionized thin shell of a spherically-symmetric nebula is unlikely to provide sufficient absorbing column to explain the radial spectral hardening. 
However, there is compelling evidence that the three-dimensional nebular morphology of \pnbd{} is an elongated ellipsoid with the long axis tilted in our direction \citep{1989ApJ...346..243M,1999MNRAS.309..731B,2005ApJ...632..340L,2016ApJS..226...15F}. 
Such a morphology implies that we view the hot bubble through more nebular material than in a spherically-symmetric case. 
Additionally, dust and molecular shells \citep{1994A&A...289..524B} and a large scattering halo comprised of ionized and neutral hydrogen gas \citep{1997AJ....113.2147H} --- also see  Figure~\ref{fig:bd30_multi} --- exist around \pnbd{}, indicating the presence of additional material that could absorb the soft X-ray emission. 

\citet{1997AJ....113.2147H} presented selective and total extinction maps of \pnbd{} using the ratio of optical and radio imaging observations. 
These maps provide clear evidence for increased absorption at the projected edge of the nebula. 
The extinction behavior is consistent with the radial spectral hardening behavior seen in the median energy image, specifically, the extinction decreases as the radial distance from the central star decreases. 
\citet{1997AJ....113.2147H} argued that a majority of the absorption is caused by the nebula itself and not the interstellar medium. 
The extinction maps of \citet{1997AJ....113.2147H} and \citet{2005ApJ...632..340L} show that the extinction is higher on the western side of the nebula than the eastern side, which is consistent with the median energy image, which is harder on the western side versus the eastern side. 
This asymmetry suggests that our spectral fit in the high median energy regions is weighted towards the eastern side of the hot bubble emission where more counts are detected. 
As a result, the range of column densities across the hot bubble are probably larger than that indicated by the best-fit values ($1.5\-3\times10^{21}~{\rm cm}^{-2}$). 

The extinction studies \citep{1997AJ....113.2147H,2005ApJ...632..340L} have argued that dust is present in the ionized nebular gas in addition to the dust and molecular shells. 
Comparing the morphology of the median energy image to that of the distribution of polycyclic aromatic hydrocarbons (PAHs) in the 3.3~$\mu$m image presented in \citet{1994A&A...289..524B} reveals a strong correspondence between the PAH emission and the radial spectral hardening morphology. 
Specifically, both images show partial ring-like morphologies with a break in similar northern locations.  
The 3.3~$\mu$m emitting region and the apparent radial spectral hardening of the X-ray plasma may be casually related. 

\section{Conclusions} \label{sec:conclusions}

Median energy imaging of the hot bubble in PN \pnbd{} provides a new perspective on the X-ray photon distribution of this well-studied planetary nebula. 
We find that the median energy image exhibits a unique morphology that is distinct from that shown in the count image. 
Specifically, a ring-like structure of higher median energy values surrounds a central region of lower median energy values that we call radial spectral hardening. 
The radial profile confirms the increase to higher median energy values towards the nebular rim. 

We explored a variety of origins to explain the observed radial spectral hardening using X-ray spectra that isolated the high-median energy and low-median energy regions. 
Three scenarios focused on the abundances, temperatures, and column densities in each region.  
We studied the behavior of the X-ray spectra under these scenarios and found that all three scenarios produced acceptable spectral fits, suggesting a degeneracy in the origin of the radial spectral hardening. 
Combinations of these scenarios, which were not considered here, could also potentially explain the radial spectral hardening.

Given that the hot bubble is embedded within a young, dense planetary nebula with an elongated nebular shell aligned with our line of sight, we suggest that the radial spectral hardening of the hot bubble X-ray emission in \pnbd{} is due to an increased column density at the projected nebular edge. 
Similar to the case of NGC 7027 \citep{2018ApJ...861...45M}, in \pnbd{} the impact of overlying nebular material on the hot bubble characteristics is important. 
In X-ray spectral analysis with existing CCD-resolution spectroscopy, assuming a constant column density across the extended emission introduces a bias that can obscure or confuse the hot bubble physical parameters. 
The median energy imaging technique provides a promising avenue for exploring such variations in relatively low count sources. 

\begin{acknowledgements}

The median energy imaging for planetary nebulae was originally developed by the author while attending the Astro Hack Week unconference in 2017 at the University of Washington and the author expresses gratitude to the organizers and participants of that workshop. 
The author is grateful to Annie Blackwell, Roel Olvera, Daniel Castro, and Regina Jorgenson who have worked with the author on applying the median energy imaging technique to supernova remnants in projects for the NSF-funded Maria Mitchell Observatory Research Experiences for Undergraduates program. 
This research has made use of data obtained from the {\it Chandra} Data Archive and software provided by the {\it Chandra} X-ray Center (CXC) in the application packages CIAO and Sherpa. 
This research made use of Astropy, a community-developed core Python package for Astronomy \citep{2018AJ....156..123A, 2013A&A...558A..33A}. 
RMJ acknowledges support from NASA contract NAS8-03060. 

\end{acknowledgements}

\vspace{5mm}
\facilities{CXO, HST(WFPC2)}

\software{CIAO \citep{2006SPIE.6270E..60F}, \texttt{astropy} \citep{2013A&A...558A..33A,2018AJ....156..123A}}

\appendix

\section{Statistical Imaging}

\begin{figure*}[ht]
\centering
\includegraphics[width=6.5in]{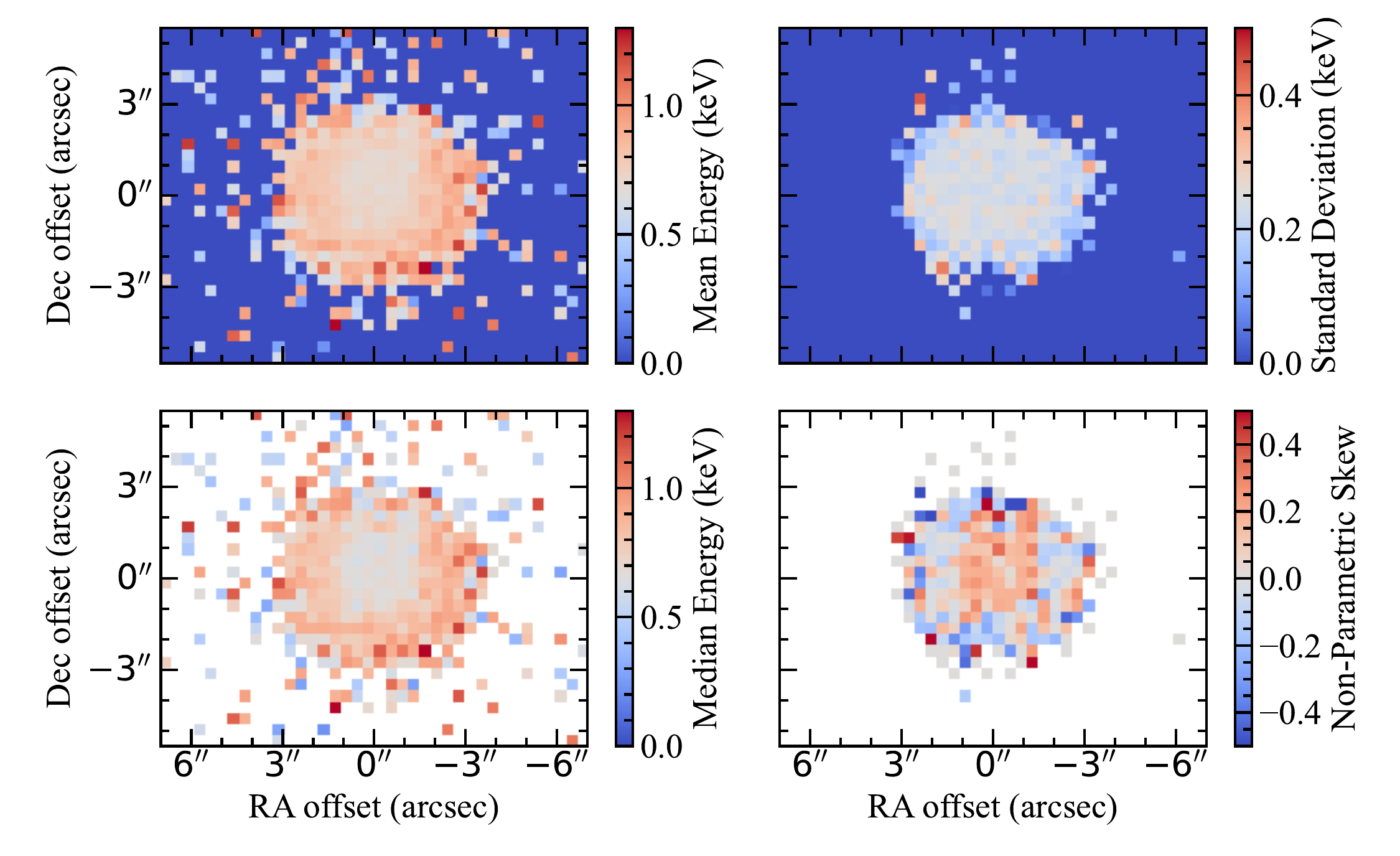}
\caption{Statistical imaging of \pnbd{}. The mean energy image, standard deviation image, median energy image, and non-parametric skew image are shown in the four panels and distinguished by the colorbar label. }
\label{fig:bd30_figa1}
\end{figure*}

In addition to the median energy image, we also considered images derived from the mean and standard deviation of the photon energy distribution in each pixel. 
From the mean image, $I_{\mu_{E}}$, the standard deviation image, $I_{\sigma_{E}}$, and median energy image, $I_{\tilde{E}}$, we calculated an image of the non-parametric skew, $I_{\rm NP}$, using 
\begin{equation}
I_{\rm NP} = \frac{I_{\mu} - I_{\tilde{E}}}{I_{\sigma_{E}}}. 
\end{equation}
The four images are shown in Figure~\ref{fig:bd30_figa1}. 

The mean energy image shares similar morphological properties as the median energy image, however, the mean energy image has less contrast between the lower and higher mean energy region compared to the median energy image. 
Note that the colorbar scales in the mean and median energy images in Figure~\ref{fig:bd30_figa1} are equal to aid in the comparison of their contrast. 
This indicates the robustness of the median value as it is less susceptible to outliers. 
The values of the standard deviation peak at $\sim0.23$~keV with the majority failing below $0.27$~keV. 
The standard deviation values appear uncorrelated with the various morphologies of the counts, mean, and median energy images. 

\begin{figure*}[ht]
\centering
\includegraphics[width=6.5in]{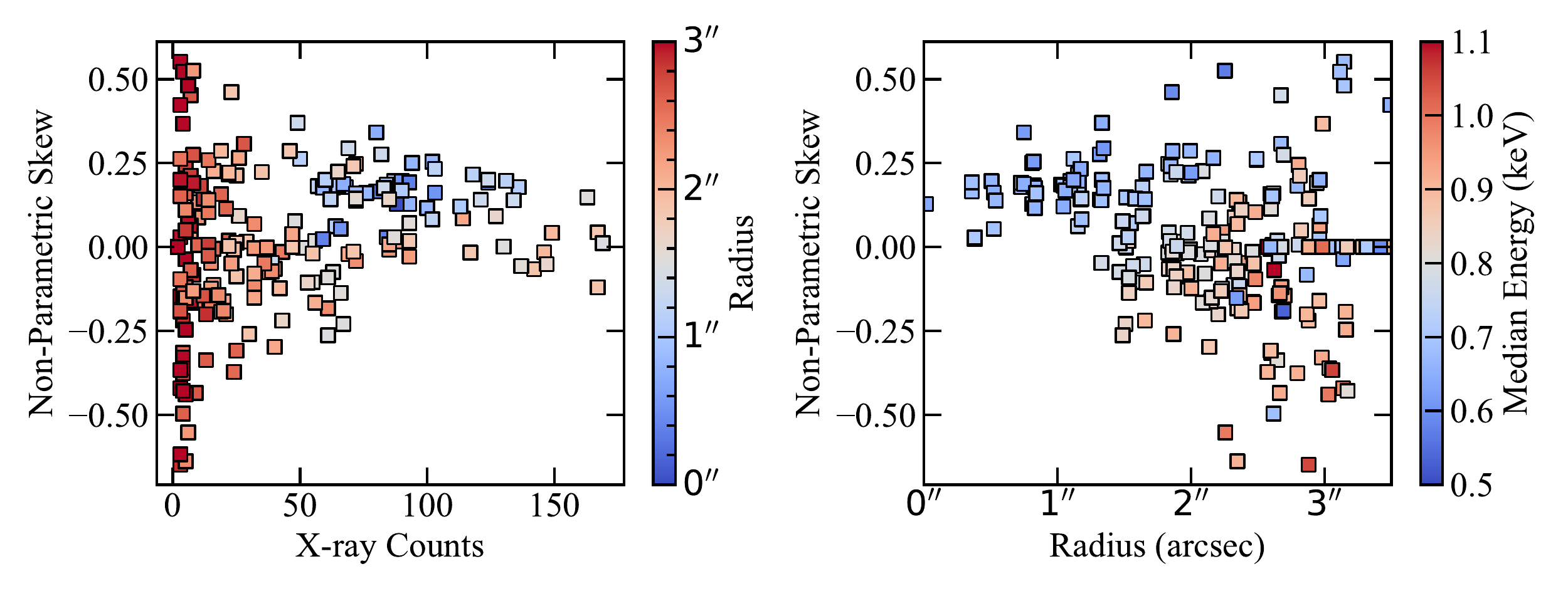}
\caption{Pixel values from the statistical images. 
In the left panel, each non-zero pixel from
the count and non-parametric skew images are plotted against each other and color-coded by the radial distance
from the pixel centroid to the central star. 
In the right panel, the radial profile of the non-parametric skew pixels
is plotted and color-coded by the median energy values. 
}
\label{fig:bd30_a2}
\end{figure*}

A strong imaging contrast arises in the non-parametric skew image with the central region showing a positive skew (mean is greater than the median) while the outer region showing a zero to slightly negative skew (mean is equal or less than median). 
We plot the pixel values in Figure~\ref{fig:bd30_a2}. 
There is no clear correlation between non-parametric skew and the number of counts. 
There is a clear indication that the central region (small radii) has a predominately positive skew, whereas above 1\farcs4 the non-parametric skew has a broader range of values. 
While the contrast of the non-parametric skew is high, perhaps more than that seen in the median energy image, it is difficult to interpret the skew since the energy distribution, i.e., the X-ray spectrum, is multi-modal. 
Nevertheless, the non-parametric skew supports the notion that the distribution of photons changes across the nebula.


\begin{thebibliography}{}

\bibitem[Akashi et al.(2007)]{2007MNRAS.375..137A} Akashi, M., Soker, N., Behar, E., et al.\ 2007, \mnras, 375, 137. doi:10.1111/j.1365-2966.2006.11267.x




\bibitem[Astropy Collaboration et al.(2018)]{2018AJ....156..123A} Astropy Collaboration, Price-Whelan, A.~M., Sip{\H{o}}cz, B.~M., et al.\ 2018, \aj, 156, 123. doi:10.3847/1538-3881/aabc4f

\bibitem[Astropy Collaboration et al.(2013)]{2013A&A...558A..33A} Astropy Collaboration, Robitaille, T.~P., Tollerud, E.~J., et al.\ 2013, \aap, 558, A33 

\bibitem[Bernard et al.(1994)]{1994A&A...289..524B} Bernard, J.~P., Giard, M., Normand, P., et al.\ 1994, \aap, 289, 524


\bibitem[Bryce \& Mellema(1999)]{1999MNRAS.309..731B} Bryce, M. \& Mellema, G.\ 1999, \mnras, 309, 731. doi:10.1046/j.1365-8711.1999.02892.x


\bibitem[Dudziak \& Walsh(1997)]{1997hstc.work..338D} Dudziak, G. \& Walsh, J.~R.\ 1997, The 1997 HST Calibration Workshop with a New Generation of Instruments, 338

\bibitem[Feigelson et al.(2005)]{2005ApJS..160..379F} Feigelson, E.~D., Getman, K., Townsley, L., et al.\ 2005, \apjs, 160, 379. doi:10.1086/432512


\bibitem[Freeman \& Kastner(2016)]{2016ApJS..226...15F} Freeman, M.~J. \& Kastner, J.~H.\ 2016, \apjs, 226, 15. doi:10.3847/0067-0049/226/2/15

\bibitem[Freeman et al.(2014)]{2014ApJ...794...99F} Freeman, M., Montez, R., Kastner, J.~H., et al.\ 2014, \apj, 794, 99. doi:10.1088/0004-637X/794/2/99

\bibitem[Fruscione et al.(2006)]{2006SPIE.6270E..60F} Fruscione, A., et al.\ 2006, \procspie, 6270



\bibitem[Harrington et al.(1997)]{1997AJ....113.2147H} Harrington, J.~P., Lame, N.~J., White, S.~M., et al.\ 1997, \aj, 113, 2147. doi:10.1086/118426

\bibitem[Heller et al.(2018)]{2018A&A...620A..98H} Heller, R., Jacob, R., Sch{\"o}nberner, D., et al.\ 2018, \aap, 620, A98. doi:10.1051/0004-6361/201832683

\bibitem[Hong et al.(2004)]{2004ApJ...614..508H} Hong, J., Schlegel, E.~M., \& Grindlay, J.~E.\ 2004, \apj, 614, 508. doi:10.1086/423445

\bibitem[Kastner et al.(2012)]{2012AJ....144...58K} Kastner, J.~H., Montez, R., Balick, B., et al.\ 2012, \aj, 144, 58. doi:10.1088/0004-6256/144/2/58

\bibitem[Kastner et al.(2008)]{2008ApJ...672..957K} Kastner, J.~H., Montez, R., Balick, B., et al.\ 2008, \apj, 672, 957. doi:10.1086/523890

\bibitem[Kastner et al.(2002)]{2002ApJ...581.1225K} Kastner, J.~H., Li, J., Vrtilek, S.~D., et al.\ 2002, \apj, 581, 1225. doi:10.1086/344363

\bibitem[Kastner et al.(2000)]{2000ApJ...545L..57K} Kastner, J.~H., Soker, N., Vrtilek, S.~D., et al.\ 2000, \apjl, 545, L57. doi:10.1086/317335


\bibitem[Kwok et al.(1978)]{1978ApJ...219L.125K} Kwok, S., Purton, C.~R., \& Fitzgerald, P.~M.\ 1978, \apjl, 219, L125. doi:10.1086/182621


\bibitem[Lee \& Kwok(2005)]{2005ApJ...632..340L} Lee, T.-H. \& Kwok, S.\ 2005, \apj, 632, 340. doi:10.1086/432875

\bibitem[Li et al.(2003)]{2003ApJ...590..586L} Li, J., Kastner, J.~H., Prigozhin, G.~Y., et al.\ 2003, \apj, 590, 586. doi:10.1086/374967

\bibitem[Li et al.(2004)]{2004ApJ...610.1204L} Li, J., Kastner, J.~H., Prigozhin, G.~Y., et al.\ 2004, \apj, 610, 1204. doi:10.1086/421866

\bibitem[Maness et al.(2003)]{2003ApJ...589..439M} Maness, H.~L., Vrtilek, S.~D., Kastner, J.~H., et al.\ 2003, \apj, 589, 439. doi:10.1086/374414

\bibitem[Marshall et al.(2004)]{2004SPIE.5165..497M} Marshall, H.~L., Tennant, A., Grant, C.~E., et al.\ 2004, \procspie, 5165, 497. doi:10.1117/12.508310

\bibitem[Masson(1989)]{1989ApJ...346..243M} Masson, C.~R.\ 1989, \apj, 346, 243. doi:10.1086/168005



\bibitem[Montez \& Kastner(2018)]{2018ApJ...861...45M} Montez, R. \& Kastner, J.~H.\ 2018, \apj, 861, 45. doi:10.3847/1538-4357/aac5df



\bibitem[Ruiz et al.(2013)]{2013ApJ...767...35R} Ruiz, N., Chu, Y.-H., Gruendl, R.~A., et al.\ 2013, \apj, 767, 35. doi:10.1088/0004-637X/767/1/35


\bibitem[Smith et al.(2001)]{2001ApJ...556L..91S} Smith, R.~K., Brickhouse, N.~S., Liedahl, D.~A., et al.\ 2001, \apjl, 556, L91. doi:10.1086/322992


\bibitem[Soker \& Kastner(2003)]{2003ApJ...583..368S} Soker, N. \& Kastner, J.~H.\ 2003, \apj, 583, 368. doi:10.1086/345343

\bibitem[Steffen et al.(2008)]{2008A&A...489..173S} Steffen, M., Sch{\"o}nberner, D., \& Warmuth, A.\ 2008, \aap, 489, 173. doi:10.1051/0004-6361:200809677

\bibitem[Toal{\'a} et al.(2019)]{2019ApJ...886...30T} Toal{\'a}, J.~A., Montez, R., \& Karovska, M.\ 2019, \apj, 886, 30. doi:10.3847/1538-4357/ab498e

\bibitem[Toal{\'a} \& Arthur(2018)]{2018MNRAS.478.1218T} Toal{\'a}, J.~A. \& Arthur, S.~J.\ 2018, \mnras, 478, 1218. doi:10.1093/mnras/sty1127

\bibitem[Toal{\'a} \& Arthur(2016)]{2016MNRAS.463.4438T} Toal{\'a}, J.~A. \& Arthur, S.~J.\ 2016, \mnras, 463, 4438. doi:10.1093/mnras/stw2307

\bibitem[Wilms et al.(2000)]{2000ApJ...542..914W} Wilms, J., Allen, A., \& McCray, R.\ 2000, \apj, 542, 914. doi:10.1086/317016

\bibitem[Yu et al.(2009)]{2009ApJ...690..440Y} Yu, Y.~S., Nordon, R., Kastner, J.~H., et al.\ 2009, \apj, 690, 440. doi:10.1088/0004-637X/690/1/440

\end{thebibliography}
\end{document}